\documentclass[iop]{emulateapj}
\usepackage{graphicx,ifthen,url,float,color,amsmath, soul,cleveref}
\newcommand{\ang}{$\mbox{\AA}$}
\newcommand{\um}{$\mu$m}
\newcommand{\ha}{\rm H$\alpha$}
\newcommand{\nii}{\rm [N{\sc ii}]}
\newcommand{\rpm}{\raisebox{.2ex}{$\scriptstyle\pm$}}
\newcommand{\vsigma}{v$_{\rm c}$/$\sigma$}

\newcommand{\kms}{\rm km\,s$^{-1}$}
\def\msol   {\ifmmode{{\rm M}_{\odot}}\else{M$_{\odot}$}\fi}
\def\mstar   {\ifmmode{{\rm M}_{\star}}\else{M$_{\star}$}\fi}
\def\vsigma {\ifmmode{{\rm v}_{\rm c}/\sigma}\else{v$_{\rm c}$/$\sigma$}\fi}
\newcommand{\lir}{L$_{\rm IR}$}

\newcommand{\lsun}{{\rm L$_{\odot}$}}
\newcommand{\hubunits}{{\rm km\,s$^{-1}$\,Mpc$^{-1}$}}

\shorttitle{\ha\ Kinematics of Three DSFGs}
\shortauthors{P.M. Drew et al.}
\begin{document} 

\title{Three Dusty Star Forming Galaxies at $z\sim1.5$: Mergers and Disks on the main sequence}
\author{Patrick M. Drew\altaffilmark{1}, Caitlin M. Casey\altaffilmark{1}, Asantha Cooray\altaffilmark{2}, and Katherine E. Whitaker\altaffilmark{3,4}}

\altaffiltext{1}{Department of Astronomy, The University of Texas at Austin, 2515 Speedway Blvd Stop C1400, Austin, TX 78712}
\altaffiltext{2}{Center for Cosmology, Department of Physics and Astronomy, University of California, Irvine, CA 92697, USA}
\altaffiltext{3}{Department of Physics, University of Connecticut, Storrs, CT 06269, USA}
\altaffiltext{4}{Department of Astronomy, University of Massachusetts, Amherst, MA 01003, USA}
\email{Email: pdrew@utexas.edu}

\label{firstpage}

\begin{abstract}

The main sequence of galaxies, a correlation between the star formation rates and stellar masses of galaxies, has been observed out to $z\sim4$. Galaxies within the scatter of the correlation are typically interpreted to be secularly evolving while galaxies with star formation rates elevated above the main sequence are interpreted to be undergoing interactions or to be Toomre-unstable disks with starbursting clumps. 
In this paper we investigate the recent merger histories of three dusty star forming galaxies, identified by their bright submillimeter emission at $z\sim1.5$.
We analyze rest-frame optical and UV imaging, rest-frame optical emission line kinematics using slit spectra obtained with MOSFIRE on Keck I, and calculate Gini and M$_{20}$ statistics for each galaxy and conclude two are merger-driven while the third is an isolated disk galaxy.
The disk galaxy lies $\sim$4$\times$ above the main sequence, one merger lies within the scatter of the main sequence, and one merger lies $\sim$4$\times$ below the main sequence. This hints that the location of a galaxy with respect to the main sequence may not be a useful discriminator of the recent star formation history of high-\mstar{} galaxies at $z\sim1$.
\end{abstract}


\section{Introduction}\label{introduction}

The majority of star forming galaxies form a correlation between their star formation rates (SFR) and stellar masses \citep[\mstar{}; e.g.][]{Rodighiero11a,Whitaker12a,Sargent14a,Schreiber15a}.
Often referred to as the main sequence of star forming galaxies (MS), the correlation has a tight scatter of $\approx$0.3\,dex in \mstar{} \citep[e.g.][]{Brinchmann04a,Noeske07a,Daddi07a,Elbaz07a,Speagle14a,Tacchella16a}.
The other, populated by starburst galaxies (SB), lies at star formation rates a few times higher than the MS at fixed \mstar{} and does not have a tight scatter \citep[e.g.][]{Rodighiero11a,Sargent14a}.
In addition to these two star forming populations, there is a population of quiescent galaxies that lie below the main sequence \citep[e.g.][]{Noeske07a,Tacchella16a,Leslie16a}.
The small scatter in the MS is typically interpreted to be the result of smooth star formation driven by the net gas inflow and outflow rate and the gas consumption rate \citep[e.g.][]{Dutton10a,Bouche10a,Tacchella16a,Scoville17a}, 
while galaxies with enhanced star formation rates are typically observed to be undergoing major mergers or shortlived starburst events (e.g. \citealt*{Mihos96a};  \citealt{Di-Matteo08a,Bournaud11a,Rodighiero11a,Whitaker12a,Silverman18a}).
Indeed, the strongest SB galaxies in the local Universe are nearly all observed to be undergoing major mergers (e.g. \citealt*{Joseph85a}; \citealt{Armus87a}; \citealt*{Sanders96a}).

Dusty star forming galaxies are a submillimeter-identified class of galaxy that are selected for their bright dust emission and characteristically elevated SFRs.
They likely peak in number density around $z\sim2$ \citep*{Casey14a} and represent a crucial phase in the evolution of $z=0$ giant elliptical galaxies \citep[e.g.][]{Swinbank04a,Swinbank06a,Engel10a,Michaowski10b,Menendez-Delmestre13a,Toft14a}. The physical cause of their bright sub-mm emission is still a matter of debate (see review by \citealt*{Casey14a}).
Those with the highest star formation rates, of order a few times 10$^{3}$\,\msol{}\,yr$^{-1}$, are virtually all driven by major mergers \citep[e.g.][]{Swinbank04a,Greve05a,Alaghband-Zadeh12a}. 
However, many studies conclude they are not unanimously merger driven, especially at more modest SFRs of $\sim$10$^2$\,\msol{}\,yr$^{-1}$ \citep[e.g.][]{Tacconi08a,Genel08a,Casey11a,Swinbank10a,Swinbank11a,Bothwell10a,Bothwell13a,Hodge12a,Drew18a,McAlpine19a}.
The selection of DSFGs is typically based on a flux density cutoff in the sub-mm, canonically $S_{\nu}$ $\gtrsim$ 2--5\,mJy, or SFR $\gtrsim$ 100\,\mstar{}\,yr$^{-1}$ for galaxies at $z\gtrsim 1$
\citep[e.g.][]{Smail97a,Barger98a,Hughes98a,Eales99a}.
This corresponds to a selection for high SFRs.
Some studies find DSFGs lie on the high-mass end of the main sequence at $z>2$ \citep{Michaowski12a,Michaowski14a} and suggest that this implies they are driven by gas accretion rather than merging \citep{Michaowski17a}.
Given their selection, for a given bin of SFR it is stellar mass that determines their location with respect to the main sequence.

With increasing redshift, the normalization of the MS increases while the typical scatter remains the same \citep[e.g.][]{Elbaz07a,Daddi07a,Rodighiero10a,Whitaker14a,Speagle14a,Scoville17a}. 
There is tension in the literature over whether mergers outside the local Universe lie above or within the MS because the SFRs of SB galaxies in the local Universe are comparable to those on the MS at moderate redshifts.
Some studies find mergers predominantly lie in the SB regime above the MS at $z>1$ \citep[e.g.][]{Kartaltepe12a,Hung13a,Cibinel19a},
or predominantly occupy higher SFRs, irrespective of the stellar mass \citep[e.g.][]{Kartaltepe10a,Ellison13a}.
Other studies suggest that the stellar masses typically measured for the elevated SB population at $z\sim2$ are unreliable and that most SB galaxies actually comprise the high-\mstar{} end of the MS \citep[e.g.][]{Michaowski12a,Michaowski14a,Koprowski14a}.

\begin{table*}
\centering
\caption{Parameters of Each Galaxy}
\begin{tabular}{c|ccccc}
\hline\hline
{\sc Source} & {\sc 450.25} & {\sc 450.27} &  {\sc 850.95} \\
\hline
RA &
10:00:28.58 & 
09:59:42.92 & 
09:59:59.80 \\ 

Dec &
+02:19:28.3 &
+02:21:45.1 & 
+02:27:07.4 \\

$z_{\rm spec}$  &
1.515 &
1.531 &
1.555 \\

M$_{\star}$ (\msol{})  &
(3.4\,{\rpm}\,0.5)$\times$10$^{11}$ &
(3.0\,{\rpm}\,0.6)$\times$10$^{11}$ &
(3.8\,{\rpm}\,3.0)$\times$10$^{10}$ \\

L$_{\rm IR}$ (\lsun) &
(1.5$^{+0.7}_{-0.5}$)$\times$10$^{12}$ &
(4.1\,{\rpm}\,0.5)$\times$10$^{12}$ &
(3.0$^{+1.2}_{-0.9}$)$\times$10$^{12}$ \\

SFR (\msol{}\,yr$^{-1}$) &
157$^{+80}_{-53}$ &
382$^{+51}_{-45}$ &
373$^{+110}_{-90}$ \\

Gini &
0.53 \rpm 0.01 &
0.63 \rpm 0.01 &
0.48 \rpm 0.01 \\

M$_{20}$ &
$-$1.725 &
$-$1.808 &
$-$1.654 \\

Physical Driver &
Merger &
Merger & 
Disk \\

\hline\hline
\end{tabular}
\label{tab:physical}

{\small {\bf Table Description} -- Positions are from \citet{Casey17a}. \mstar\ is estimated using MAGPHYS \citep{da-Cunha08a} with the HIGHZ extension \citep{da-Cunha15a}. The errors on \mstar\ differ slightly from \citet{Casey17a} because they are estimated using a range of SFHs from a continuous star formation history to an instantaneous burst, following the procedure of \citet{Hainline11a}. The errors on \lir\ and, as a direct result, SFR are estimated following the procedure of \citet{casey12a}.}
\end{table*}

The determination of galaxy classification is vital to addressing the question of what role mergers play in galaxy evolution through cosmic time. Numerous techniques have been employed to determine galaxy classification including imaging, kinematics, and non-parametric analyses (see review by \citealt{Conselice14a}).
Imaging studies classify galaxies based on visual signatures of mergers, interactions, or disks in images \citep[e.g.][]{Hubble26a,deVaucouleurs63a,Abraham96a,Abraham96b,Brinchmann98a,Kartaltepe15a}.
While large imaging studies make this kind of analysis relatively easy to perform, they may fall short in a few key ways at $z\gtrsim1$ when trying to distinguish mergers from disks.
At these redshifts, the optical waveband begins to probe rest-frame UV emission originating from younger stars. Not only is this light more highly dust obscured than at longer rest-frame wavelengths, but the UV is not a sensitive probe of the bulk of the stellar mass, which most closely resembles the distribution of the total mass of the galaxy.
Additionally, kinematically regular galaxies may appear morphologically disturbed in imaging \citep[e.g.][]{Bournaud08a}, because galaxies at $z>1$ tend to be clumpier than their low-$z$ counterparts \citep[e.g.][]{Abraham96a,Elmegreen06a}. 
This may lead to disks being misclassified as mergers or irregular galaxies.
Also, surface brightness dimming may hide features of a merger, such as tidal tails (e.g. \citealt*{Hibbard97a}).

Kinematic studies overcome some of the issues associated with imaging studies. They measure the motions of gas inside galaxies via observations of the doppler shift of emission lines.
Kinematic observations of disk galaxies exhibit smooth rotational fields, while mergers show more complex velocity fields (see \citealt{Glazebrook13a} for a review).
Kinematic studies are still susceptible to misclassification however, especially if observed near or shortly after coalescence or if observed at low spatial resolution \citep[e.g.][]{Hung15a}.

Non-parametric analyses \citep[e.g.][]{Abraham03a,Lotz04a,Lotz08a} rely on the grouping of galaxies in parameter space to identify galaxy morphology and assembly history. The strength of this technique is that it can be applied to any type of galaxy without prior knowledge about the form the model should take. Its weakness is that the distinction between classes may not be as clear as with imaging or kinematic studies \citep{Conselice14a}.

In this paper we present rest-frame UV and rest-frame optical imaging, a rest-frame optical emission line kinematic analysis, and a non-parametric analysis of three DSFGs at $z\sim1.5$ as case studies of high SFR-selected galaxies.
Our goal is to identify their physical drivers.
We then compare their location in the SFR-\mstar{} plane with their physical drivers to investigate whether merging DSFGs lie above, within, or below the MS.

Section \ref{sec:obs} of this paper describes our observations and data reduction, Section \ref{sec:results} presents our imaging and kinematic analyses, Section \ref{sec:Gini-M20} describes Gini-M$_{20}$ statistics, Section \ref{sec:MS} discusses the galaxies in the context of the MS, and section \ref{sec:conc} summarizes.
Throughout this work we adopt a {\it Planck} $\Lambda$CDM cosmology with H$_{0} = 67.7$\,\hubunits, $\Omega_{\Lambda} = 0.6911$ \citep{Planck-Collaboration16a} and a Chabrier initial mass function \citep[IMF;][]{Chabrier03a}.

\section{Observations}\label{sec:obs}

\begin{figure*}
    \centering\includegraphics[width=0.99\textwidth]{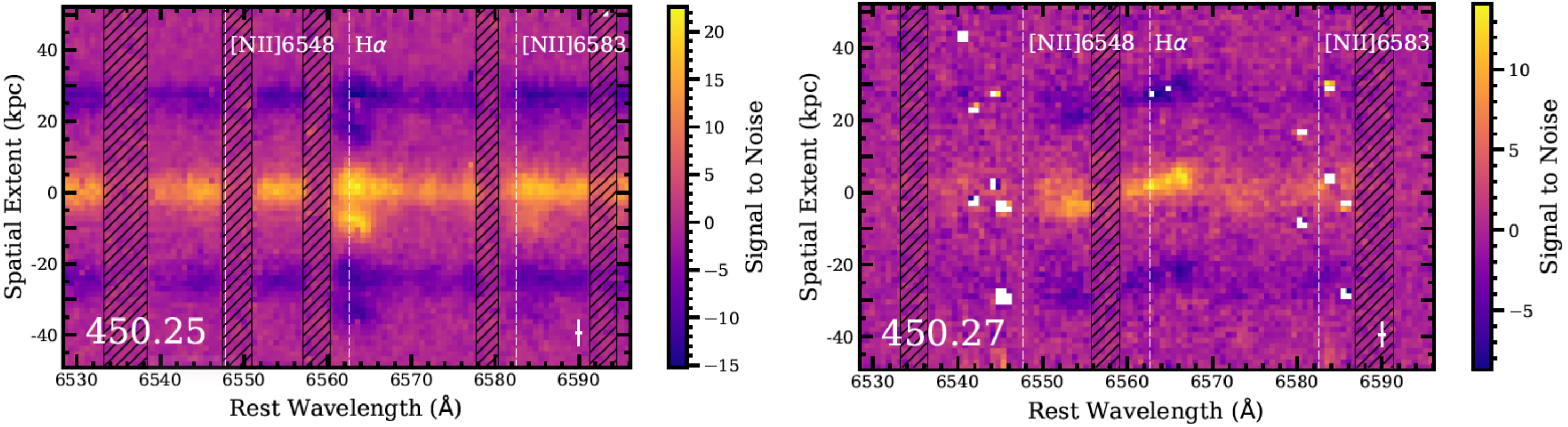}
    \caption{Signal to noise spectra for 450.25 and 450.27 showing \ha, [NII], and continuum emission. Hatched regions denote telluric line contamination and white regions denote bad pixels. The resolution in each dimension is denoted by the white cross in the lower right corners. Regions of negative SNR are artifacts from the ABBA nod pattern coaddition procedure.}
    \label{fig:SNRs}
\end{figure*}

The spectroscopic data presented in this manuscript were obtained with the Multi-Object Spectrometer For Infra-Red Exploration \citep[MOSFIRE;][]{Mclean10a,McLean12a} on Keck I as part of a spectroscopic follow-up campaign to measure the redshifts of DSFGs identified at flux densities $>$12.4\,mJy and $>$2.4\,mJy at 450\,\um\ and 850\,\um\ respectively with {\sc Scuba-2} in the COSMOS field \citep[see][]{Casey13a,Casey17a}.
The galaxies in the present paper were selected from their parent sample based on their high signal to noise (SNR) and spatially resolved \ha\ and [NII] emission.
The three galaxies presented in this paper are the only ones that have sufficiently resolved emission to allow for a kinematic analysis from the original 114 targets in the \citet{Casey17a} sample.
One galaxy, named 850.95, is also published in \citet[][hereafter D18]{Drew18a} as an observational counter example to the hypothesis that galaxies at intermediate redshifts may have declining rotation curves.
The rotation curve of this galaxy is flat in the outer galaxy, much like typical disk galaxy rotation curves at $z=0$. In this paper we discuss the \ha\ kinematics of 850.95 and refer the reader to D18 for a discussion of its dark matter content.

The three galaxies presented in this paper, 450.25, 450.27, and 850.95, have prefixes of either 450 or 850, corresponding to the wavelength at which they were initially identified in \citet{Casey13a}. Only 450.27 is detected at both 450\,\um\ and 850\,\um.
\autoref{tab:physical} lists basic characteristics of each galaxy.
For additional details about the parent sample and its selection, see \citet{Casey17a}.

H-band spectroscopic observations of 450.25, 450.27, and 850.95 were obtained on 2013 December 31 at W. M. Keck Observatory. The full width at half maximum (FWHM) of seeing was 0\farcs85. Galaxy 450.25 was observed for a total integration time of 2880\,s, 450.27 for 1320\,s, and 850.95 for 1920\,s. The slit width was set to 0\farcs7 and a 1\farcs5 ABBA nod pattern was used between exposures. 
The spectra were reduced with the MOSPY Data Reduction Pipeline\footnote{\href{http://keck-datareductionpipelines.github.io/MosfireDRP/}{http://keck-datareductionpipelines.github.io/MosfireDRP/}}, and one-dimensional spectra were extracted using the {\sc iraf}\footnote{{\sc iraf} is distributed by the National Optical Astronomy Observatories, which are operated by the Association of Universities for Research in Astronomy, Inc., under cooperative agreement with the National Science Foundation.} package, {\sc apall}.
Apertures of extraction were placed on each pixel with an aperture radius of half the average seeing. Adjusting the aperture size does not significantly change the radial velocity and velocity dispersion measurements presented in the following sections. Variance weighting was used in the spectral extraction.

\autoref{fig:SNRs} shows SNR spectra for 450.25 and 450.27, including \ha, [NII], and continuum emission. See figure 1 in \citet{Drew18a} for the SNR spectrum of 850.95. The white crosses denote the seeing in the vertical dimension and the combined instrument resolution and seeing in the horizontal dimension. The negative images to the north and south of the galaxy are characteristic of the data coaddition step associated with the nod pattern. Slits were randomly oriented with respect to the galaxies because MOSFIRE slit masks prevent custom orientations for individual galaxies when observing in multiplex mode.

\begin{figure}
    \centering\includegraphics[width=0.99\columnwidth]{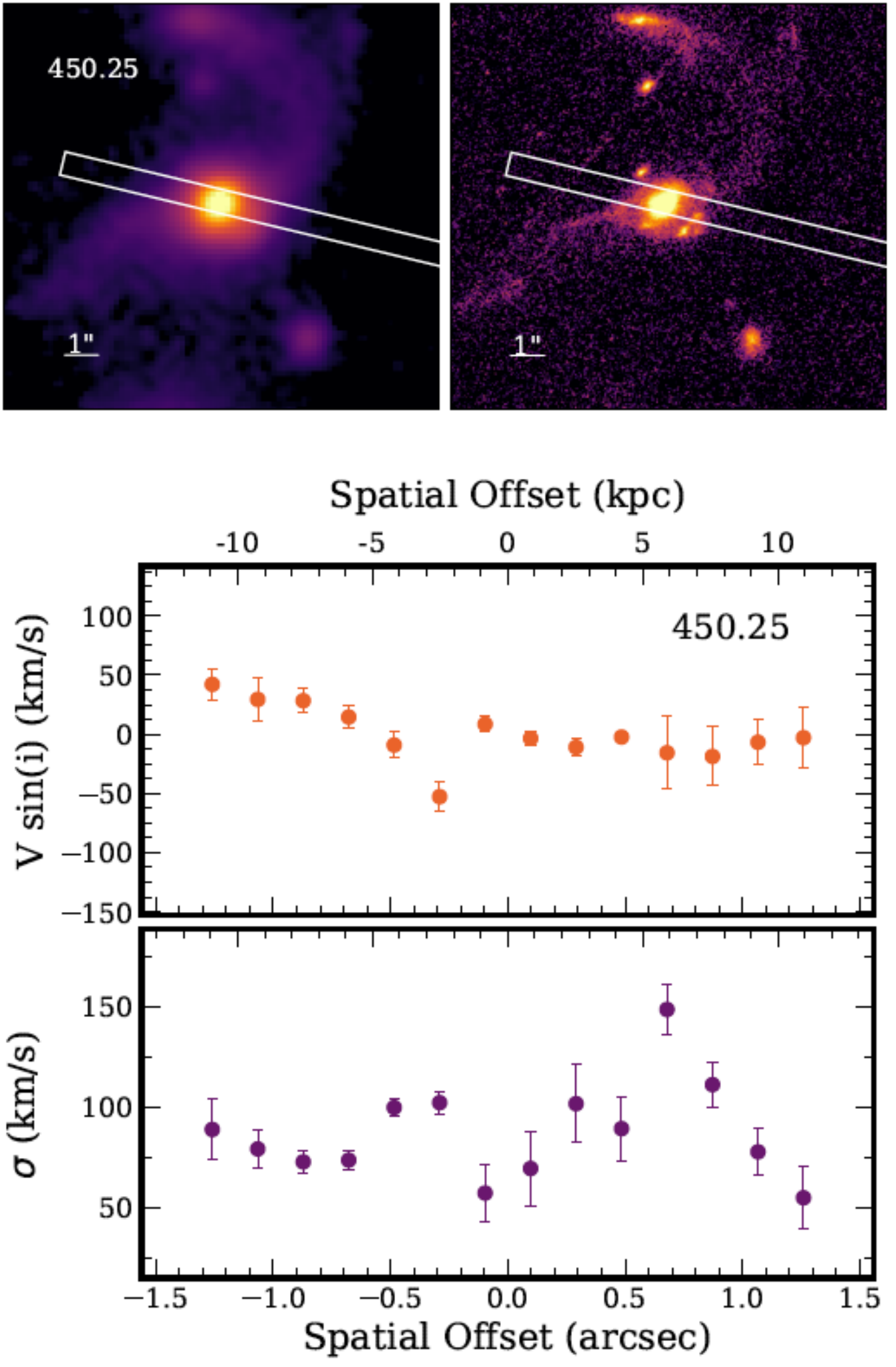}
    \label{fig:image45025}
    \caption{The top panels show ground-based H-band (rest-frame $\sim$6400\,\ang; \citealt{McCracken12a}) and {\it HST} F814W (rest-frame $\sim$3200\,\ang; \citealt{Koekemoer07a}) imaging on the left and right, respectively, with the MOSFIRE slit overplotted in white. The tidal tails seen in both images suggest 450.25 is undergoing an interaction or is in the early stages of a merger. The bottom panels show position-velocity and position-dispersion diagrams measured along the MOSFIRE slit. They are disordered fields showing no symmetry.}
    \label{fig:PV45025}
\end{figure}

\begin{figure}
    \centering\includegraphics[width=0.99\columnwidth]{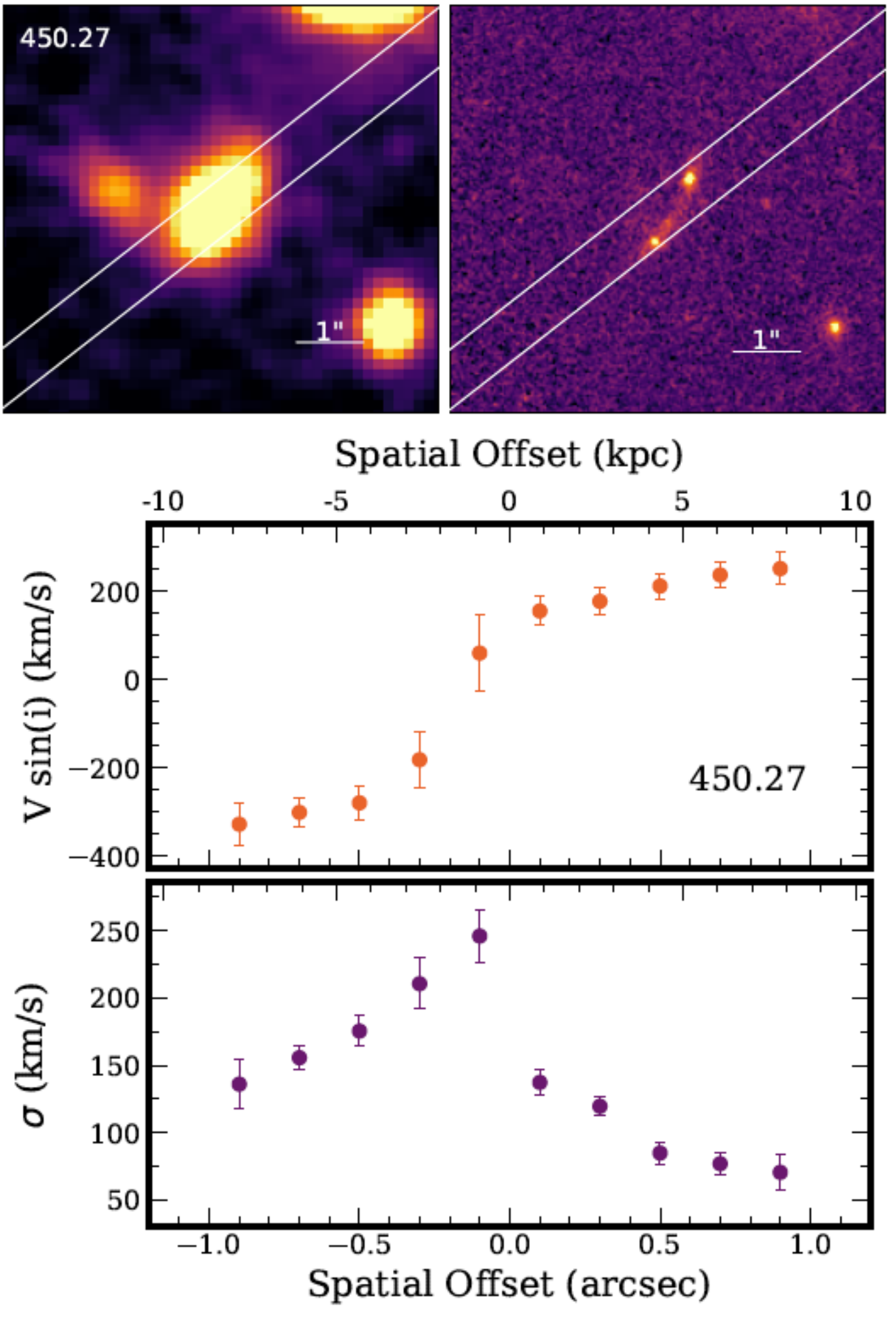}
    \caption{The top panels show ground-based H-band (rest-frame $\sim$6400\,\ang) and {\it HST} F814W (rest-frame $\sim$3200\,\ang) imaging on the left and right, respectively, with the MOSFIRE slit overplotted in white.
    The continuum emission running perpendicular to and outside of the MOSFIRE slit in the H band image likely belongs to a galaxy at higher redshift not physically associated with 450.27 because the colors of the two systems are very different. The bottom panels show position-velocity and position-dispersion diagrams measured along the MOSFIRE slit.
    The velocity curve looks Keplerian but the dispersion curve is not symmetric in the magnitudes of velocities measured on each side of the galaxy.}
  \label{fig:PV45027}
\end{figure}

We simultaneously fit Gaussians to \ha, \nii$\lambda$6548 and \nii$\lambda$6583 to the extracted apertures using the {\sc iraf} package, {\sc splot}, forcing the centroids to have fixed spacing and the Gaussian widths to be tied. We exclude pixels contaminated with telluric emission in the fits. Errors in fit centroids and widths are derived using 1000 Monte Carlo perturbations of the data by sky noise. The bottom panels of figures \ref{fig:PV45025} and \ref{fig:PV45027} show the position-velocity and position-dispersion diagrams measured from the MOSFIRE spectra. These will be discussed further in Section \ref{sec:results}. We find no evidence of broad line emission in any of the spectra and \citet{Casey17a} finds no evidence of x-ray emission or mid-IR SED slopes in these galaxies that would indicate luminous active galactic nuclei are present.

The imaging includes H band from the UltraVISTA survey (rest-frame $\sim$6400\,\ang; \citealt{McCracken12a}) and \textit{Hubble} F814W (rest-frame $\sim$3200\,\ang; \citealt{Koekemoer07a}) data. 
While Y, J, H, and Ks imaging is available for these galaxies from the UltraVISTA survey, we present the H band because it matches the band the spectra are observed in. The H band imaging is consistent with the longest available from the UltraVISTA survey, the Ks band. These observations are the longest-wavelength high spatial resolution imaging available.
The F814W imaging is the longest-wavelength \textit{HST} imaging available.
The images are presented in the top rows of Figures \ref{fig:PV45025} and \ref{fig:PV45027} and will be discussed further in Section \ref{sec:results}.

\section{Kinematics and Morphologies}\label{sec:results}

We classify our galaxies based on their position-velocity and position-dispersion diagrams into two categories: mergers or disks.
In the position-velocity diagram we expect a disk to have smooth, symmetric velocity field about a single spatial axis. In the position-dispersion diagram we expect a peak centered on the spatial symmetry axis of the position-velocity diagram.
Evidence for a merger would be a disrupted position-velocity and position-dispersion fields and/or a discontinuity in velocity between two galaxies \citep[e.g.][and references therein]{Glazebrook13a}.

\subsection{450.25}
The top row of \autoref{fig:PV45025} shows H-band (left; rest-frame $\sim$6400\,\ang) and {\it Hubble} filter F814W imaging (right; rest-frame $\sim$3200\,\ang) of 450.25.
Tidal tails connect two galaxies and indicate this is a merging system.
The straight-line separation between the two main galaxy components is of order 50\,kpc, indicating this is an early stage merger or interaction prior to coalescence.
The H-band emission is not well fit by a S\'ersic profile \citep{Sersic63a}. There are significant warps in the fit residual map caused by the merger. 
The H-band S\'ersic fitting process is discussed further in Section \ref{sec:Gini-M20}.
The classification of merger is determined without the need to consider the position-velocity and position-dispersion diagrams, however we present them here for completeness.
The bottom two panels of \autoref{fig:PV45025} show the position-velocity and position-dispersion diagrams of 450.25 measured from the MOSFIRE spectrum.
The velocity and dispersion fields show disorder characteristic of a merger. We do not correct the measured velocities for inclination or spatial resolution effects.

\subsection{450.27}
\label{subsec:450.27}
\begin{figure}
    \centering
    \includegraphics[width=0.99\columnwidth]{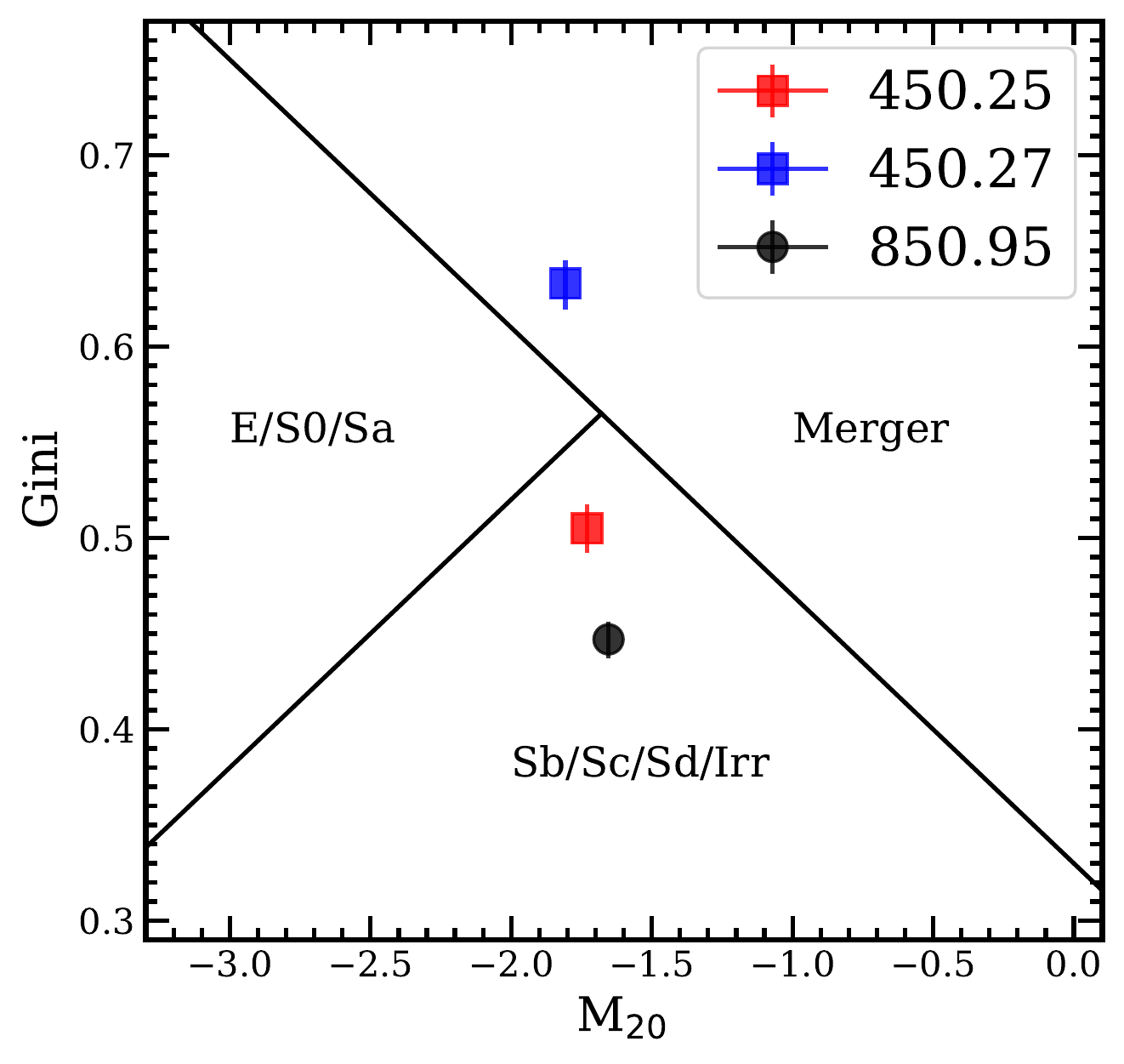}
    \caption{Gini-M$_{20}$ diagram with our sample overplotted. Squares indicate classifications of mergers, while the circle indicates disk, as determined by our morphological and kinematic analyses. The black boundary lines are adopted from \citet{Lotz08a}, which are calibrated to galaxies between $0.2<z<1.2$ in the EGS \textit{HST} survey. While 450.25 is undergoing a merger it falls in the disk region. This may be due to the fact that it is observed in an early stage of the merger and H-band emission is therefore not yet concentrated in a small region.
    The imaging and kinematic analysis of 450.27 was a little ambiguous (see Section \ref{subsec:450.27}), but given its location in the Gini-M$_{20}$ diagram we conclude it is a merger.
    Errors on Gini, calculated by performing 10$^4$ statistical bootstrap measurements of the pixels associated with each galaxy. Errors on M$_{20}$ were not performed.}
    \label{fig:Gini-M20}
\end{figure}

The top row of \autoref{fig:PV45027} shows H-band (left; rest-frame $\sim$6400\,\ang) and {\it HST} filter F814W (right; rest-frame $\sim$3200\,\ang) imaging of 450.27.
The H-band image shows a perpendicular emission component approximately 1$\arcsec$ to the west of the targeted component that is not visible in the rest-frame UV. 
We conclude this is an unassociated galaxy at higher redshift based on the very different photometric colors between the components, though confirmation of this would require follow-up observations.
The \textit{HST} image shows two distinct star forming knots with emission centroids separated by $\sim$9\,kpc.
These knots may be two star forming knots in a single galaxy or they may be two separate galaxies close to coalescence. The knots are each well-fit by S\'ersic profiles with bright central cores, suggesting perhaps that they could be two separate galaxies close to coalescence. 
The northern knot seen in the upper right panel of \autoref{fig:PV45027} has a S\'ersic index of $n=1.2$ with a half light radius of $r_{1/2}=1.9$\,kpc and the southern knot has a S\'ersic index of $n=0.6$ with a half light radius of $r_{1/2}=2.8$\,kpc.
However, the rest-frame UV is expected to be $>$90\% extincted in galaxies with star formation rates as high as 450.27 \citep{Whitaker17a}.
The H-band emission is not well fit by a S\'ersic profile because the best-fit parameters are unphysical. Its residual map shows two peaks coincident spatially with the UV knots.
The H-band S\'ersic fitting process is discussed further in Section \ref{sec:Gini-M20}.

The bottom two panels of \autoref{fig:PV45027} show position-velocity and position-dispersion diagrams for 450.27. The total range in radial velocities is $\sim$600\,\kms, which is a lower limit because we do not correct for galaxy inclination.
The velocity curve looks Keplerian. It is smoothly varying and symmetric. The velocity dispersion on the other hand looks disturbed. It is asymmetric in the magnitude of velocities on either side of the galaxy. This may be a signature of a galaxy merger or interaction.
Recent works by \citet{Hung15a} and \citet{Simons19a} demonstrate that mergers at high redshift may display disk-like kinematics when observed at low spatial resolution.
As we will discuss in Section \ref{sec:Gini-M20}, this galaxy lies in the merger region of the Gini-M$_{20}$ diagram. Taken together with the imaging and kinematics, we conclude this is likely a merging system.

\subsection{850.95}
The third galaxy we analyze is 850.95, which is first presented by D18. Figure 2 of D18 shows the galaxy in H-band (left; rest-frame $\sim$6400\,\ang) and \textit{Hubble} F814W imaging (right; rest-frame $\sim$3200\,\ang). The H-band emission profile is best fit by an exponential disk with a S\'ersic index of $n=1.29\rpm0.03$ and a disk inclination of $i=87\rpm2^{\circ}$. The right panel of Figure 2 in D18 shows an offset between the rest-frame UV and dust continuum emission, possibly as a result of the near edge-on orientation of the disk. Figure 4 in D18 shows position-velocity and position-dispersion diagrams of 850.95 with clear kinematic signatures of ordered disk rotation. The position-velocity diagram shows a smooth, symmetric velocity field with a flat outer-galaxy rotation velocity of $285\rpm12$\,\kms. The curve is well-fit by an arctangent function. The position-dispersion diagram shows a smooth, symmetric, centrally peaked dispersion field with a systemic ionized gas velocity dispersion of $48\rpm4$\,\kms. The exponential disk emission profile along with the smooth arctangent velocity profile strongly suggest 850.95 is an example of a DSFG that is not undergoing a major merger.

\section{Gini and M$_{20}$ Diagnostics}\label{sec:Gini-M20}
Next we compute the non-parametric diagnostics Gini and M$_{20}$ (e.g. \citealt*{Abraham03a}, \citealt{Lotz04a}, \citealt{Lotz08a}) for each of our galaxies. 
The Gini statistic quantifies how the light is distributed throughout a galaxy. 
A Gini value of 1 would imply all the emission originates from a single pixel, while a Gini value of 0 would imply a uniform light distribution across multiple pixels.
M$_{20}$ measures the second moment of the galaxy's brightest 20\% of pixels relative to the total second moment.
These two quantities have been shown to roughly separate mergers, ellipticals, and disk galaxies \citep[e.g.][]{Lotz08a}.

In order to perform the analysis, first we isolate the pixels associated with each galaxy above a threshold SNR of 8 in H-band imaging using the Python package, Photutils \citep{Bradley19a}. Thresholds are chosen to be as low as possible while still separating unrelated field galaxies in the segmentation maps.
Next, we run the Photutils source deblending routine on galaxy 450.27 in order to separate the two perpendicular components seen in the H-band imaging (see \autoref{fig:PV45027}). Whether or not this deblending is performed does not change the Gini-M$_{20}$ classification of merger for this galaxy.
Finally we use the Python package, Statmorph \citep{Rodriguez-Gomez19a} to measure Gini and M$_{20}$ statistics from the segmentation maps.
Statmorph also simultaneously fits S\'ersic profiles to the pixels associated with each galaxy in the segmentation maps, the results of which we discussed in previous sections.

\autoref{fig:Gini-M20} shows our galaxies in Gini-M$_{20}$ space. Galaxy 450.27 lies within the merger region, while galaxies 450.25 and 850.95 lie within the disk region. Our kinematic and morphological analyses conclude 450.25 is a merger, 850.95 is a disk, and 450.27 is possibly a merger, although it is a bit ambiguous. Considering this together with the location of 450.27 in the Gini-M$_{20}$ diagram, we conclude that it is indeed a merger.
The Gini-M$_{20}$ plot in Figure 7 from \citet{Lotz08a} shows that the overwhelming majority of galaxies that lie in the merger region of Gini-M$_{20}$ space are true mergers.
This figure also shows that mergers may lie in any region of Gini-M$_{20}$ space.
Galaxy 450.25 is confirmed via imaging to be undergoing a merger or interaction  (see Figure \ref{fig:PV45025}), but it lies within the disk region of the Gini-M$_{20}$ diagram.
This may be caused by the fact that it is in an early merger/interaction stage so the H-band emission is not yet concentrated in a just few pixels.

\section{Main Sequence of Galaxies}\label{sec:MS}

\begin{figure}
    \centering
        \includegraphics[width=0.99\columnwidth]{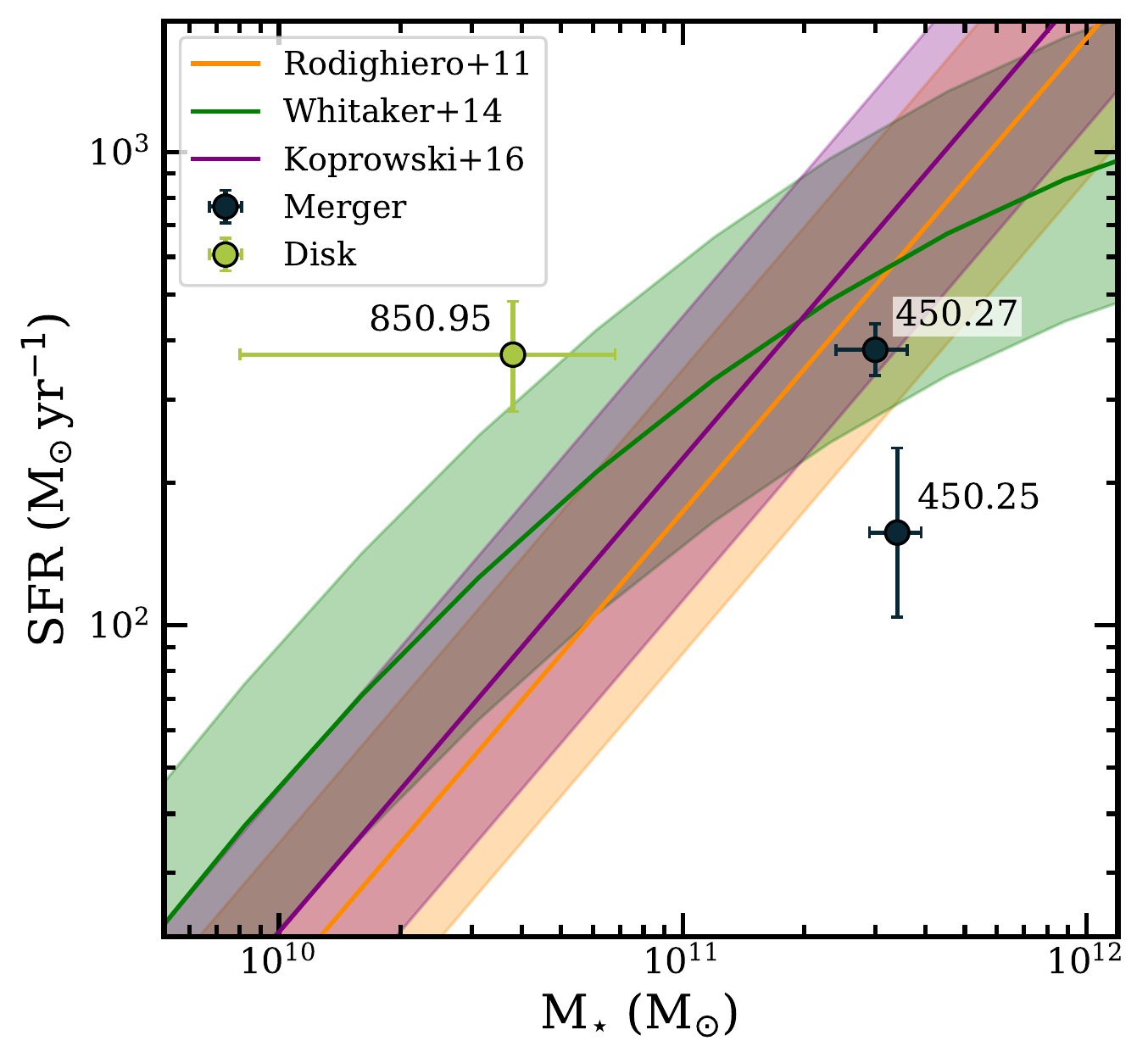}
        \caption{Our sample plotted against three main sequence fits from the literature. Our galaxies comprise the high stellar mass end of the MS. The orange line is the fit from \citet{Rodighiero11a} to data between $1.5<z<2.5$, the green line is the fit from \citet{Whitaker14a} to data between $1.5<z<2.0$, and the purple line is the fit from \citet{Koprowski16a} to data at $z>1.5$. The shaded regions are $\rpm$0.3\,dex from each fit, which is the typical 1$\sigma$ scatter observed in the distribution \citep[e.g.][]{Whitaker14a}.
        }
    \label{fig:MS}
\end{figure}

Now we consider the merger/disk classifications in the context of the main sequence of star forming galaxies.
The star formation rates and stellar masses of our sample are reported by \citet{Casey17a}, who measure these quantities using the high-$z$ extension of MAGPHYS \citep{da-Cunha08a,da-Cunha15a} using multi-wavelength photometry (UV through sub-mm) from the COSMOS collaboration \citep{Capak07a,Laigle16a}.
In the present paper, to account for systematic uncertainties on stellar mass caused by the assumption of a star formation history, we follow the procedure of \citet{Hainline11a}, which was developed to estimate uncertainties on \mstar{} for similarly-selected DSFGs at $z\sim2$.
To summarize, we take the errors on \mstar{} to be half the difference between the stellar masses estimated using instantaneous burst histories and those estimated using continuous star formation histories.
Works by \citet{Michaowski12a,Michaowski14a} demonstrate that different assumed SFHs can strongly affect the derived \mstar{}.
This systematic uncertainty from choice of SFH is larger than those reported by the MAGPHYS fits.

\autoref{fig:MS} shows our sample plotted on the main sequence of star forming galaxies at $z\sim1.5$ from three works in the literature.
The first \citep{Rodighiero11a} is fitted to far- and near-IR selected galaxies between $1.5<z<2.5$ in the COSMOS and GOODS-South fields using combined UV and IR SFRs.
The second \citep{Whitaker14a} is fitted to galaxies from the 3D-HST photometric catalogs between $1.5<z<2.0$ using combined UV and IR SFRs.
The third \citep{Koprowski16a} is fitted to 850\,$\mu$m selected galaxies at $z>1.5$ from the S2CLS survey \citep{Geach17a} using UV through sub-mm SEDs.

Galaxy 450.27 is the only galaxy to lie within the 0.3\,dex scatter of the main sequence. Galaxy 850.95 is the only disk galaxy in our sample but it lies at a SFR $\sim4 \times$ greater than the MS. Galaxy 450.25 is a merger with visible tidal tails but it lies at a SFR $\sim4 \times$ less than the MS.
The overall distribution of the galaxies is consistent with the works of \citet{Michaowski12a,Michaowski14a,Michaowski17a} and \citet{Koprowski16a}, who find that DSFGs roughly comprise the high \mstar{} end of the MS. 
In particular, all three galaxies fit within the scatter of the samples at $1.0<z<1.5$ and $1.5<z<2.0$ in Figure 10 of \citet{Michaowski17a}.

\citet{Michaowski17a} suggest because DSFGs comprise the high \mstar{} end of the MS major mergers are not a dominant driver of their star formation rates.
Mergers are short-lived events that are expected to elevate the star formation rates of galaxies above the main sequence.
The present paper, as well as many other works \citep[e.g.][]{Di-Matteo08a,Hung13a,Cibinel19a}, demonstrate that merging systems may lie on the main sequence.
\citet{Puglisi19a} find that up to 50\% of the most massive galaxies on the MS at $z\sim1.3$ may have star formation driven by merging.
Major merger activity may not enhance star formation rates at every stage of a merger (e.g. \citealt*{Bergvall03a}; \citealt{Di-Matteo08a,Jogee09a,Narayanan15a,Fensch17a,Silva18a}).
Additionally, high resolution hydrodynamical simulations by \citet{Fensch17a}, show that mergers at high redshift have lower star formation efficiencies compared with those at low redshift.
We caution the reader that a detailed analysis of the merger classifications of DSFGs on the high \mstar{} end of the MS needs to be performed before it can be concluded that they are not undergoing merging.

It is interesting that the SFR of 450.25, a clear early stage merger, is $\lesssim$4$\times$ the MS value at its \mstar{}. 
Quiescence is known to correlate with galaxy compactness \citep[e.g.][]{Bell12a,Lee18a}, but the location of 450.25 in Gini-M$_{20}$ space demonstrates the galaxy is not compact. 
It is possible that 450.25 was previously quenched but the merger has yet to fully turn star formation on again. A study of the molecular gas content of this galaxy would be illuminative.

\section{Summary}\label{sec:conc}
Mergers or galaxy interactions drive the star formation rates of galaxies lying at higher SFRs than the main sequence at $z=0$ \citep[e.g.][]{Mihos96a}, however at higher redshift where the star formation rates of galaxies at all \mstar{} are elevated, the distinction between these two populations is less clear.
Kinematic observations of DSFGs in the literature reveal a mix between merging and secularly evolving galaxies \citep[e.g.][]{Swinbank06a,Alaghband-Zadeh12a} but such studies are limited to small sample sizes because they are observationally expensive.
The determination of where DSFGs lie in the SFR-\mstar{} plane is important to help determine their importance in galaxy evolution through the cosmos.
Some studies find that DSFGs sit above the main sequence \citep[e.g.][]{Hainline11a}, but recent work by \citet{Michaowski12a,Michaowski14a,Michaowski17a} has determined the average DSFG actually comprises the high stellar mass end of the MS, and conclude that this is evidence that major mergers do not drive their star formation rates.

In this paper we combine imaging, kinematic, and non-parametric analyses to determine whether three sub-mm identified DSFGs at $z\sim1.5$ have merger or secular recent SHFs.
We find two to be undergoing merging or interactions and one to be an isolated disk galaxy.

Rest-frame UV and optical imaging of galaxy 450.25 shows tidal tails, which are clear evidence it is undergoing a merger or interaction.
The rest-frame optical emission is not well fit by a S\'ersic profile because the best-fit parameter values are unphysical. This is due to warps caused by the interaction/merger.
Position-velocity and position-dispersion diagrams reveal disorder characteristic of a merger. 
The Gini coefficient and M$_{20}$ values, $G=0.534\rpm0.006$ and M$_{20} = -1.725$, place 450.25 in the disk region of the Gini-M$_{20}$ diagram, possibly because this is an early stage merger or interaction.

Rest-frame UV imaging of galaxy 450.27 shows two distinct star forming knots which we conclude are likely two galaxy cores close to coalescence. Rest-frame optical imaging shows one distinct emission region rather than two knots, although it is observed through seeing with a FWHM comparable to the separation between the two UV knots.
The rest-frame optical emission is not well fit by a S\'ersic profile because the best-fit parameter values are unphysical.
The position-velocity diagram looks Keplerian while the position-dispersion diagram looks asymmetric in the magnitude of velocity dispersion on each side of the galaxy.
The Gini coefficient and M$_{20}$ values, $G=0.633^{+0.007}_{-0.005}$ and M$_{20} = -1.808$, place 450.27 in the merger region of the Gini-M$_{20}$ diagram.

Rest-frame optical imaging of galaxy 850.95, published by D18, is well fit by an exponential disk profile with S\'ersic index $n=1.29\rpm0.03$. The rest-frame UV emission is clumpy and is offset from dust continuum emission detected with ALMA.
The position-velocity and position-dispersion diagrams show clear signatures of rotation with a velocity profile well-fit by an arctangent function and a centrally peaked, symmetric dispersion curve.
The Gini coefficient and M$_{20}$ values, $G=0.481\rpm0.005$ and M$_{20} = -1.654$, place 850.95 in the disk region of the Gini-M$_{20}$ diagram.

Despite its disk classification, 850.95 sits at a SFR $\sim$4$\times$ above the MS, placing it in the starburst regime. Despite merging activity, 450.27 lies within the scatter of the MS, and 450.25 lies $\sim$4$\times$ below the MS. It is unexpected that a merging galaxy has a lower star formation rate than typical galaxies on the main sequence. Further investigation as to the cause of the suppressed star formation is needed.

Our sample hints
that perhaps the specific star formation rate (SFR/\mstar{}) is not a useful discriminator of the recent merger history of high-\mstar{} galaxies at $z>1$.
A detailed investigation of a statistical sample of DSFGs is needed in order to determine the recent star formation histories of DSFGs both on and off the main sequence in order to determine the physics driving their star formation rates.

\acknowledgements
The authors thank Justin Spilker and Jorge Zavala for useful discussions, as well as the anonymous
reviewer who provided valuable comments and suggestions.
P.D. acknowledges financial support by a NASA Keck PI Data Awards: 2012B-U039M, 2011B-H251M, 2012B-N114M, 2017A-N136M, NSF grants AST-1714528 and AST-1814034, and the University of Texas at Austin College of Natural Sciences.
The authors wish to recognize and acknowledge the very significant cultural role and reverence that the summit of Mauna Kea has always had within the indigenous Hawaiian community. We are most fortunate to have the opportunity to conduct observations from this mountain.
Some of the data presented herein were obtained at the W.M. Keck Observatory, which is operated as a scientific partnership among the California Institute of Technology, the University of California and the National Aeronautics and Space Administration. The Observatory was made possible by the generous financial support of the W.M. Keck Foundation.
This research made use of APLpy, an open-source plotting package for Python hosted at http://aplpy.github.com, Astropy, a community-developed core Python package for Astronomy \citep{Astropy13a}, and the python packages Matplotlib \citep{Matplotlib07a}, Numpy \citep{Numpy11a}, and Pandas \citep{McKinney10a}.

\end{document}